\documentclass[reprint,amsmath,amssymb,aps,prl]{revtex4-1}

\usepackage{graphicx}
\usepackage{dcolumn}
\usepackage{bm}
\usepackage{hyperref}
\usepackage{xcolor}

\newcommand{\df}{\; \mathrm{d}}
\newcommand{\eps}{\varepsilon}

\newcommand{\red}[1]{#1}
\newcommand{\minor}[1]{#1}  
\newcommand{\del}[1]{}
\newcommand{\delm}[1]{}

\newcommand{\ilm}{Univ Lyon, Univ Claude Bernard Lyon 1, CNRS, Institut Lumi\`ere Mati\`ere, F-69622, VILLEURBANNE, France}

\begin{document}


\title{Giant thermoelectric response of nanofluidic systems \red{driven by water excess enthalpy}}

\author{Li Fu}
\affiliation{\ilm}
\author{Laurent Joly}%
\email{laurent.joly@univ-lyon1.fr}
\affiliation{\ilm}
\author{Samy Merabia}%
\email{samy.merabia@univ-lyon1.fr}
\affiliation{\ilm}

\date{\today}

\begin{abstract}
Nanofluidic systems could in principle be used to produce electricity from waste heat, but current theoretical descriptions predict a rather poor performance as compared to thermoelectric solid materials. 
Here we investigate the thermoelectric response of NaCl and NaI solutions confined between charged walls, using molecular dynamics simulations. 
We \red{compute} 
a giant thermoelectric response, two orders of magnitude larger than the predictions of standard models. 
We show that water excess enthalpy -- neglected in the standard picture -- plays a dominant role in combination with the electroosmotic mobility of the liquid-solid interface. Accordingly, the thermoelectric response can be boosted using surfaces with large hydrodynamic slip. 
Overall, the heat harvesting performance of the model systems considered here is comparable to that of the best thermoelectric materials, and the fundamental insight provided by \minor{molecular dynamics} suggests guidelines to further optimize the performance, opening the way to recycle waste heat using nanofluidic devices. 
\end{abstract}

\maketitle


\paragraph{Introduction}
With a fast-growing energy consumption, and energy production based mostly on fossil fuels, our society is in crucial need of new, sustainable energies. Nanofluidic systems could play a key role in the development of such new energies \cite{Sparreboom2009,Bocquet2014}. For instance, blue energy systems based on membranes with nanoscale porosity can produce electricity from salinity difference with very good efficiency, opening the way to large-scale harvesting of the osmotic energy of sea water \cite{Siria2013,Feng2016,Siria:2017kt}.
At the core of nanofluidic energy conversion systems lie the so-called electrokinetic (EK) effects, coupling different types of transport in nanochannels \cite{Anderson1989,Bocquet2010}.
In aqueous electrolytes, EK effects arise from the dynamics of the electrical double layer (EDL), a diffuse layer of non-neutral liquid in the vicinity of charged surfaces, whose thickness -- the Debye length $\lambda_\text{D}$ -- is typically nanometric in aqueous electrolytes \cite{Andelman1995,IsraelachviliBook,LyklemaBook}. 

EK energy conversion has been studied since the early 60s, but it has found a renewed interest with the advent of nanofluidic systems, offering significant efficiency improvements \cite{VanderHeyden2006,Pennathur2007}, as predicted theoretically \cite{Pagonabarraga2010,Hartkamp2018} 
and measured experimentally \cite{xue2017water}. 
Hydroelectric energy conversion has been studied extensively in the last 15 years 
\cite{Siria:2017kt,Bonhomme2017,Blanc2018,Zhao:2013ji,xie2011strong}. 
The increased performance arises from several mechanisms specific to the nanoscale 
\cite{Liu2018}, e.g. liquid-solid slip \cite{Sparreboom2009,Ajdari:2006ci,Ren2008,Rankin2016}. 
However, the possibility to use nanofluidic systems as thermoelectric converters has only been discussed very recently \cite{Bonetti2015,Huang2015,barragan2018perspectives,Salez2018,Kristiansen2019,Li2019}. 
Traditional thermoelectric semiconductors offer high thermoelectric performance at room temperature, but their use is limited owing to their toxicity and rarity \cite{barragan2018perspectives}. With that regard, EK effects have been studied and some analytical models have been suggested \cite{barragan2018perspectives,Dietzel:2016fy,Dietzel:2017co,DiLecce:2017hy,DiLecce2018, ly2018nanoscale,li2018thermoelectricity,zhang2018temperature}. 
In particular, a recent theoretical study reported enhanced Seebeck coefficient in confined electrolyte solutions \cite{Dietzel:2016fy,Dietzel:2017co}, an effect explained by the electrostatics of the EDL, in the spirit of a standard picture developed by Derjaguin and collaborators \cite{Derjaguin1980,Derjaguin:1987bf}. 

Here, we perform molecular dynamics (MD) simulations to explore the physical mechanisms at play in the thermoelectricity of nanofluidic channels\minor{, and we} 
show that the thermoelectric response of confined electrolytes can be orders of magnitude higher than what is predicted by standard models\minor{.} \delm{based on the electrostatics of the EDL. This strongly enhanced response is to a large extent driven by the water excess enthalpy, which is usually neglected in the standard picture. As a consequence, we demonstrate that nanofluidic systems could be a potential alternative to traditional thermoelectric solutions based on solid-state materials.}

\begin{figure}
	\centering
	\includegraphics[width=\linewidth]{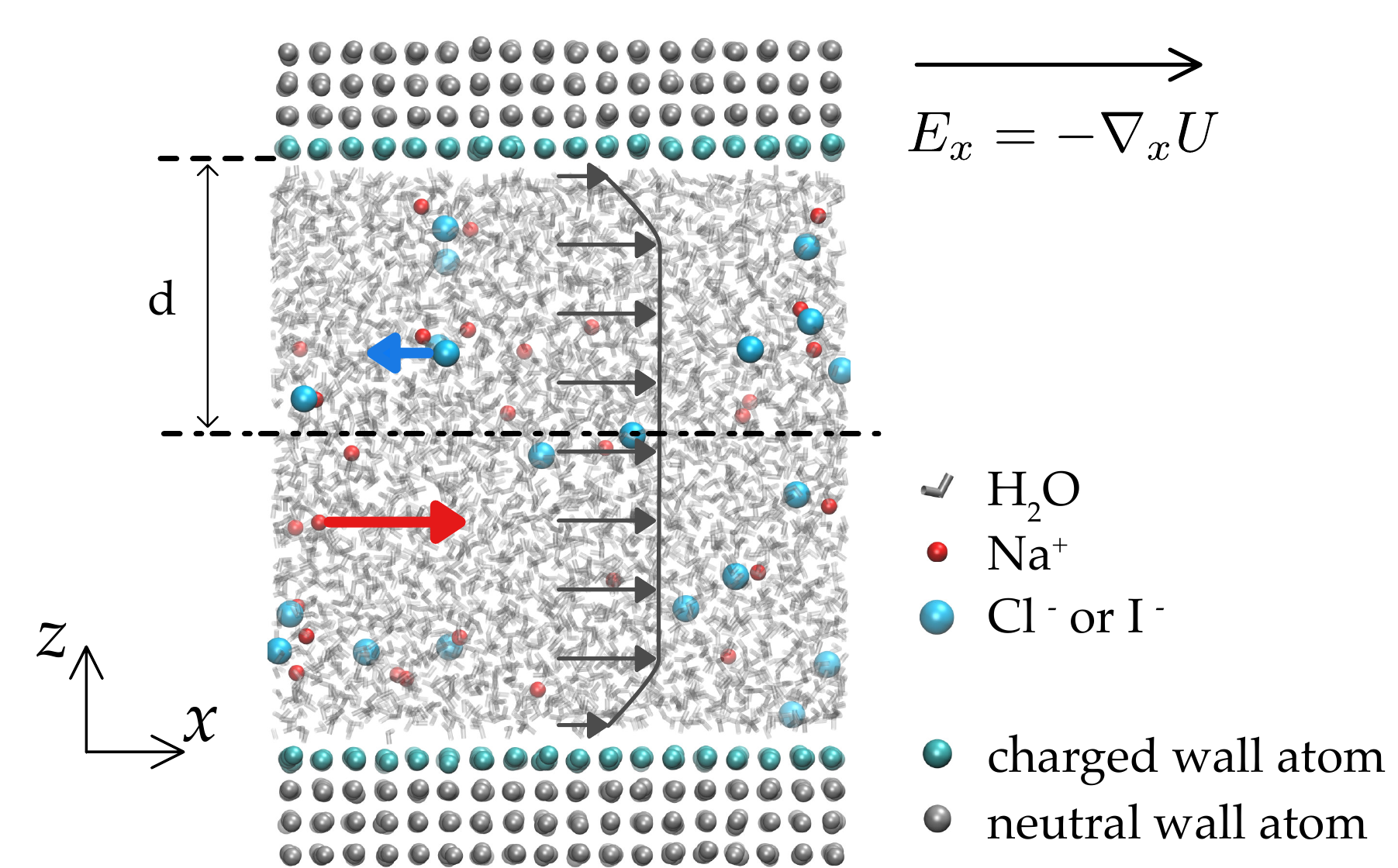}
	\caption{Illustration of the simulation system. An aqueous electrolyte solution is confined in a slit nano-channel with charged inner surfaces. An external electric field is applied and the heat flux is \red{computed}. 
	The arrows indicate the directions of ion motion.} 
	\label{fig:Illustration_systems}
\end{figure}

\paragraph{Methods}
Under external gradients of electric potential $-\nabla U$ and of temperature $-\nabla T$, the thermoelectric response of a fluidic system can be described by the non-diagonal terms of the response matrix \cite{deGroot:1984ue}:
\begin{equation}
\left[ \begin{array}{c} j_e \\ j_h \end{array} \right] = \begin{bmatrix} \sigma & M_{12} \\ M_{21} & \kappa T \end{bmatrix} \times \left[ \begin{array}{c} -\nabla U \\ -\nabla T/T \end{array} \right],
\label{eq:coefficient_definition}
\end{equation}
where $j_e$ is the electric current density, $j_h$ is the heat flux density, $\sigma$ and $\kappa$ are the electrical and thermal conductivities of the system, and $M_{ij}$ are phenomenological coefficients. 
$M_{12}$ characterizes the so-called Seebeck effect -- the conversion of heat into electricity,  
and $M_{21}$ describes the Peltier effect -- generation of an excess heat flux in an electric field. According to Onsager reciprocal relations, $M_{12} = M_{21} = M_{te}$ \cite{deGroot:1984ue, Brunet:2004bv}. In the following, we will refer to $M_{te}$ as the thermoelectric coefficient.

We conducted MD simulations with \delm{the} LAMMPS \delm{package} \cite{Plimpton1995} to \red{compute} 
the thermoelectric response of aqueous \red{electrolyte} solutions 
confined between two parallel Einstein solid walls (Fig.~\ref{fig:Illustration_systems}) and explored, in particular, the effect of surface charge. 
\red{We considered two electrolytes, NaCl and NaI, which were shown to display 
different electro-osmotic responses due to ionic specificity \cite{Huang2007,Huang2008}.}
We present here the main features of the simulations, and we reported details in the Supplemental Material \cite{sm}.

We used the TIP4P/2005 model~\cite{Abascal:2005ka} for water, and the scaled-ionic-charge model by \citet{Kann:2014kd} for ions. 
\red{Liquid-solid interactions were taken from a previous MD study \cite{Huang2007,Huang2008} of electro-osmosis on a generic hydrophobic surface (contact angle $\sim 140\,^\circ$), unless specified. We will come back on the large value of the contact angle in the following.} 
\red{Beyond the value of the surface charge, we explored the role of charge distribution by considering homogeneously or heterogeneously charged walls \cite{sm}. 
Counter-ions were added in the liquid to ensure electroneutrality.} 

The bulk electrolyte concentration $\rho_b$ was set to ca. $0.35$\,M unless specified \red{(we will also present simulations without salt)}. 
\red{This large salt concentration ensures that the Debye length, $\lambda_\text{D} \approx 5$\,\AA{} is $\sim 10$ times smaller than the system height, so that the EDLs do not overlap.} 

To \red{obtain} 
the thermoelectric coefficient, we maintained the system at $T = 298$\,K \red{(applying a Nos\'e-Hoover thermostat to the liquid, only on the degrees of freedom perpendicular to the flow)} and $p = 1$\,atm \cite{sm}, 
we applied different external electric fields $E_x = -\nabla_x U$ between 0.05 and 0.2\,V/nm in independent simulations, and we computed the excess heat flux density $j_h$ as detailed in previous work \cite{Fu:2017fr,Fu:2018cz} and in Ref. \cite{sm}: 
\begin{equation}
	\label{eq:enthalpy_flux}
	j_h = \frac{1}{2d} \int_{-d}^d \delta h(z) v(z) \df z,
\end{equation}  
where $d$ is the half height of channel, $\delta h$ is the excess enthalpy density and $v$ is the velocity. 
The linearity of the response to the electric fields was checked and the linear regression of $j_h$ against $E_x$ gave the thermoelectric coefficient $M_{te}=M_{21}=j_h/E_x$. 
\red{Note that the large electric fields considered here are commonly used in MD studies to extract the induced fluxes from thermal noise \cite{Huang2007,Huang2008,Joly2014,Yoshida2014,Predota2016,Freund02,Qiao2004}; indeed, it has been shown that electrokinetic response coefficients obtained with such electric fields were consistent with equilibrium results obtained through the linear response theory, i.e., in the limit of vanishing forcing \cite{Yoshida2014,Marry03bis,Dufreche05bis}.}  
%

\begin{figure}
	\centering
	\includegraphics[width=\linewidth]{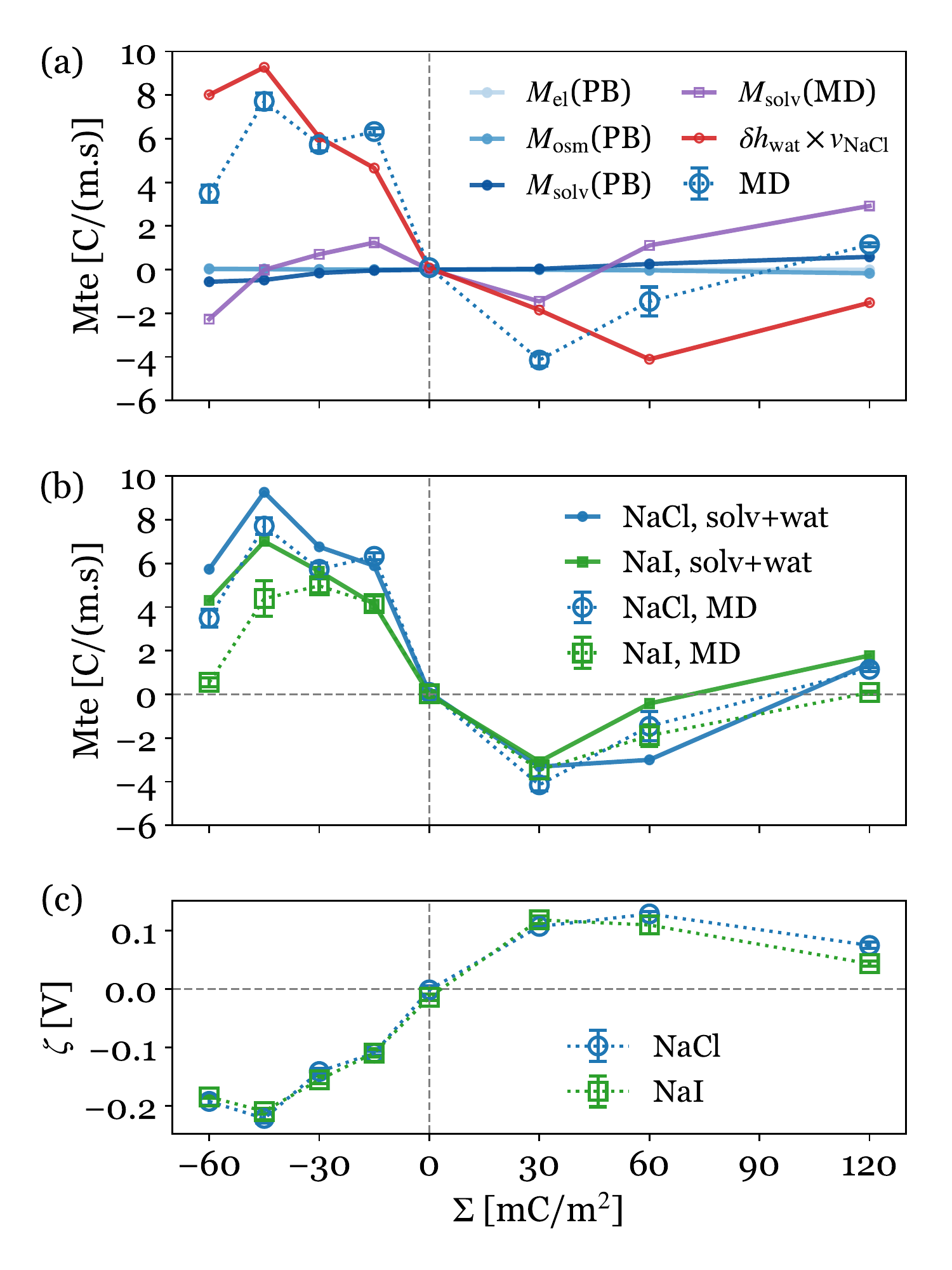}
	\caption{(a)~Thermoelectric coefficient $M_{\rm te}$ as a function of the surface charge $\Sigma$, for NaCl solutions with homogeneously charged hydrophobic surfaces. Open symbols: MD results. Solid lines: theoretical estimates of different contributions (see text for detail). 
	(b)~$M_{\rm te}$ versus $\Sigma$, comparison between NaCl and NaI. Open symbols: MD results.  Solid lines: \red{model} 
	taking into account the water and solvation contributions (see text for detail). 
	(c)~\red{Computed} 
	$\zeta$ potential as a function of $\Sigma$ for NaCl and NaI solutions.
	}
	\label{fig:Mte_sigma}
\end{figure}

\paragraph{Results and discussion}
Fig.~\ref{fig:Mte_sigma}~(a-b) displays the evolution of the thermoelectric coefficient $M_{\rm te}$ on a homogeneously charged hydrophobic surface as a function of the surface charge. 
For the two salts, $|M_{\rm te}|$ goes through a maximum on both positively and negatively charged surfaces, and the dependency of $M_{\rm te}$ against $\Sigma$ is highly asymmetrical. 
Also, the response can be quite different for NaCl and NaI, see e.g. $\Sigma = -60$\,mC/m$^2$ in Fig.~\ref{fig:Mte_sigma}~(b). All these observations are in contrast with Derjaguin's treatment of thermoelectricity in charged liquids, which predicts $M_{\rm te} \propto \Sigma^3$ \cite{Derjaguin1980,Derjaguin:1987bf}.

We will now try to capture these numerical results theoretically. 
Our starting point is the expression of the excess enthalpy density $\delta h(z)$ in Eq.~\eqref{eq:enthalpy_flux}, which for electrolytes can be decomposed in two terms:
\begin{equation}
	\label{eq:excess_enthalpy}
	\delta h(z)=\delta h_{\rm w}(z) + \delta h_{\rm EDL}(z) .
\end{equation}  
The first term corresponds to the excess enthalpy of water molecules, and is commonly neglected in treatments of thermoelectricity \cite{Derjaguin1980,Derjaguin:1987bf}, and the second term is related to the EDL. This latter contribution can be decomposed into three terms:
\begin{equation}
	\label{eq:excess_enthalpy_EDL}
	\delta h_{\rm EDL}(z)=\delta h_{\rm el}(z) + \delta h_{\rm osm}(z) + \delta h_{\rm solv}(z) .
\end{equation}  
The first term, $\delta h_{\rm el}$, has an electrostatic origin, and is the one considered in the standard picture; the second term, $\delta h_{\rm osm}$, originates from the osmotic pressure of the counter-ions, and the last term, $\delta h_{\rm solv}$, arises from the solvation enthalpy of the ions. The expression of the electrostatic and osmotic excess enthalpy is given in Ref. \cite{sm}, 
and in the following we concentrate on the solvation term, as it will be shown to be larger than the two other EDL contributions.  In the limit of separate EDLs, the solvation excess enthalpy  writes: 
\begin{equation}
  \label{eq:solvatation_enthalpy}
  \delta h_{\rm solv}(z) = h^+ \left\{ n^+(z)-n^+_b \right\} +h^- \left\{ n^-(z)-n^-_b \right\} ,
\end{equation}
where $n^\pm(z)$ denote the ion number density, $n^\pm_b$ their bulk value, and $h^{\pm}$ the ion solvation enthalpy. Here we used the bulk values of $h^{\pm}$ to estimate the solvation term of all the ions. 
Figure~\ref{fig:Mte_sigma}~(a) shows the three EDL contributions to the thermoelectric coefficient for NaCl solutions. 
The electrostatic part and the osmotic part are smaller than the simulation results, by two and one order of magnitude, respectively. This implies that the thermoelectric response 
of the confined electrolyte is much larger than what is predicted by the standard picture. 
The solvation term is much larger than the other EDL contributions, but still smaller in amplitude than the \red{computed} 
$M_{te}$. 
Let us now estimate the water contribution in Eq.~\eqref{eq:excess_enthalpy}. To that aim, we used the enthalpy profile as computed in additional simulations on a system of pure confined water, keeping all the other parameters constant. 
In order to \red{calculate} 
the resulting coefficient $M_{\rm te}$, we convoluted this approximate excess enthalpy with the electroosmotic velocity \red{obtained from} 
the simulations. 
Figure~\ref{fig:Mte_sigma}~(a) shows that the water contribution is dominant over the EDL terms, and is of the same order of magnitude as the NaCl simulation results. The water term provides a good description of the simulated $M_{\rm te}$ for negatively charged surfaces. However, it tends to overestimate $\vert M_{\rm te} \vert$ for positively charged surfaces and also for negative highly charged surfaces. Hence, the water term does not capture all the complexity of the thermoelectric response of the confined liquid.

To confirm this statement, we have added to $\delta h_{\rm w}$ the solvation term $\delta h_{\rm solv}$ and compared to the simulation results for both NaCl and NaI in Fig.~\ref{fig:Mte_sigma}~(b). The corresponding solvation term is calculated beyond the Poisson-Boltzmann approximation, by directly using the \red{simulated} 
ionic density profile. For NaCl, the agreement is very good over the range of $\Sigma$ considered, while for NaI the agreement is only partial. This partial agreement may be explained by the fact that both the water and solvation terms are approximate, in so far as we used bulk enthalpies to describe the solvation of confined ions, and the water term is calculated for a pure water system.  Nevertheless, from Fig.~\ref{fig:Mte_sigma}~(a-b) we conclude that the thermoelectric response of the confined electrolyte is dominated by the water contribution, and to a lesser extent by the solvation contribution.

Apart from the excess enthalpy density, the thermoelectric response is intimately related to the electroosmotic (EO) mobility, 
usually quantified in terms of the so-called zeta-potential \red{through the Helmholtz-Smoluchowski (HS) equation} \cite{Delgado2007,Hartkamp2018}: 
$v_{\rm EO} = -\frac{\eps}{\eta} \zeta \times E_x$,
with $\eps$ and $\eta$ the bulk permittivity and shear viscosity, respectively. 
The EO mobility can be amplified by liquid-solid slip \cite{Marry03bis,Joly2004,Bocquet2010,Hartkamp2018}, which is quantified by the slip length $b$ (the extrapolated depth where the no-slip boundary condition would apply). \red{This amplification has been studied theoretically \cite{Muller1986,Stone2004,Joly2004,Joly2006} and experimentally \cite{Bouzigues2008,Audry2010}, showing that} 
the zeta-potential writes $\zeta = \phi_0 + \Sigma b / \eps$, with $\phi_0$ the surface potential. 
\red{We estimated $\zeta$ following the experimental procedure, by computing the EO velocity under an electric field and applying the HS equation \cite{sm}.} 
Figure \ref{fig:Mte_sigma}~(c) shows \red{its} 
non-monotonous behavior 
as a function of the surface charge, which is due to the decrease of the slip length $b$ with $\Sigma$ \cite{Joly2006,Huang2008,Botan2013,Jing2015}. This complex behavior of $\zeta$ in combination with the excess enthalpy density profile, results in the non-monotonous and asymmetrical behavior of $M_{\rm te}$. 

Through the EO mobility, the thermoelectric response can also be enhanced by slip. 
To illustrate this point, we focused on NaCl and a homogeneous charge of $-30$\,mC/m$^2$, and we 
\red{tuned liquid-solid slip by considering} surfaces with different wetting properties: one hydrophilic ($\theta \sim 60\,^\circ$) with a low slip length of ca. $0.2$\,nm, and another very hydrophobic ($\theta \sim 180\,^\circ$) with a large slip length of ca. $6.8$\,nm \red{(details on contact angle estimation can be found in Ref. \cite{sm})}. 
Although the \red{original hydrophobic and the very hydrophobic surfaces} 
might appear unrealistic in terms of wetting, the values of slip lengths we obtained are \red{measured experimentally on moderately hydrophobic surfaces \cite{Bocquet2010}, and even larger slip lengths have been observed} 
on new 2D materials or in nanotubes \cite{Majumder2005,Holt2006,Maali2008,Secchi2016}. 
\red{Moreover, 
the amplitude of the interfacial enthalpy excess did not change much when we increased the contact angle from $\sim 60\,^\circ$ to almost $180\,^\circ$, so that the thermoelectric responses computed in this work -- controlled by both the interfacial enthalpy excess and liquid-solid slip -- should not be unrealistic.} 
For the low-slip surface we obtained a very small value of $M_{\rm te} = -0.32 \pm 0.05$\,C/(m.s). 
In contrast, we obtained a very large value of $M_{\rm te} = 21.45 \pm 0.54$\,C/(m.s) on the high-slip surface. 
Liquid-solid slip therefore represent a powerful lever to optimize thermoelectric conversion in nanofluidic systems. 

All the results discussed so far concern systems characterized by a uniform surface charge distribution\red{, mimicking e.g. a polarized surface}. 
\red{However, surface charge can also result from randomly distributed charged groups, e.g. on silica. Therefore,}  
we also simulated heterogeneously charged surfaces,  
and for both NaCl and NaI solutions, the values of $M_{\rm te}$ were smaller by typically a factor of $10$ \cite{sm}. 
Note that the excess enthalpy density profiles remained similar between homogeneous and heterogeneous surfaces, and that the decrease of $M_{\rm te}$ can be mostly related to different hydrodynamic boundary conditions \cite{sm}. 

Finally, we focus on the energy harvesting applications of such nanofluidic systems. We evaluate the performance of thermoelectric energy conversion with nanofluidic devices, by computing their Seebeck coefficient $S_e$, and their so-called figure of merit denoted $ZT$, traditionally used to quantify the performance of thermoelectric materials. 
The Seebeck coefficient is defined as $S_e = -\nabla V / \nabla T$ when $j_e = 0$. It then results from Eq.~\eqref{eq:coefficient_definition} that $S_e = M_{te} /(\sigma T)$. 
$ZT$ is expressed as a function of the Seebeck coefficient $S_e$, the thermal conductivity, the electric conductivity and the temperature: $ZT = \sigma S_e^2 T/ \kappa$. The figure of merit can equivalently be expressed as a function of the thermoelectric coefficient $M_{\rm te}$: $ZT = M_{\rm te}^2 / (\sigma \kappa T)$. 
To quantify the \red{expected experimental} figure of merit, we assume in the following that the solid walls may be chosen in order to have limited influence on the device thermoelectric response. 
In particular, \red{we use the experimental thermal conductivity} 
of water, $\kappa_{\rm water} = 0.609$\,W/m/K, and we assume that the solid walls are electric insulators with a negligible thermoelectric response, so that the electric conductivity $\sigma$ and thermoelectric coefficient $M_{\rm te}$ are those of the confined liquid, \red{computed} 
in the simulations. 

\begin{figure}
	\centering
	\includegraphics[width=0.96\linewidth]{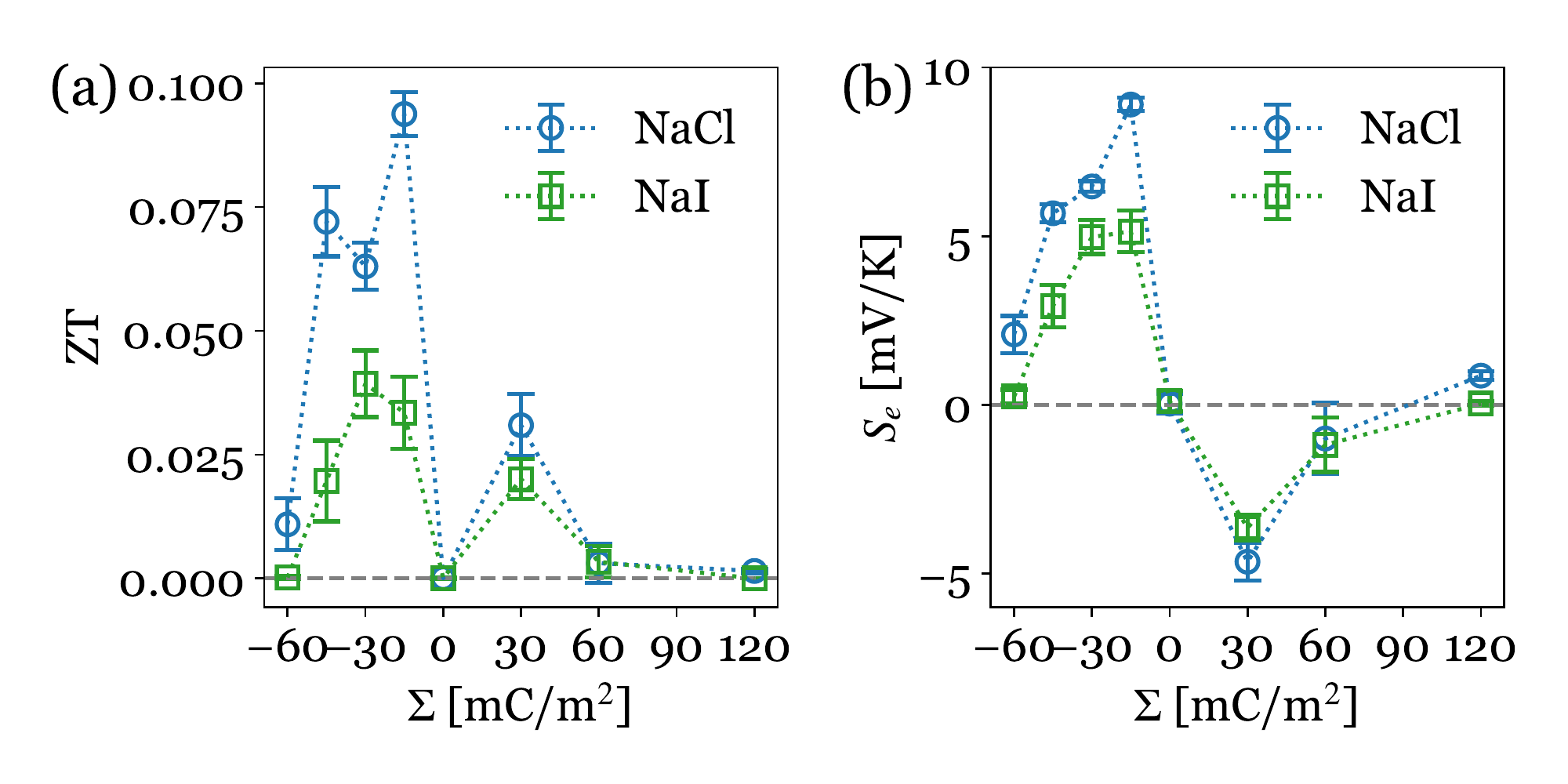}
	\caption{(a)~Figure of merit $ZT$ and (b)~corresponding Seebeck coefficient $S_e$ against the surface charge density $\Sigma$, for the same systems studied in Fig.~\ref{fig:Mte_sigma}.}
	\label{fig:ZT_Seebeck_sigma}
\end{figure}

Figure~\ref{fig:ZT_Seebeck_sigma} displays the computed $ZT$ and $S_e$ against the surface charge density, for the same systems studied in Fig.~\ref{fig:Mte_sigma}. 
NaCl generally offers better performance than NaI. A maximum $ZT$ of ca. $0.1$ is obtained for $\Sigma = -15$\,mC/m$^2$, corresponding to a Seebeck coefficient of ca. $10$\,mV/K. 
In a recent theoretical work \cite{Dietzel:2016fy}, $|S_e|$ originating from the electrostatic term was estimated to be ca. $0.2$\,mV/K in a confined fluidic system, with a similar $\zeta$ potential and the same ratio of the slit gap to the Debye length. 
\red{Other simulation studies evaluated the Seebeck coefficient of bulk electrolyte solutions \cite{DiLecce:2017hy,DiLecce2018}, and found absolute values of up to $\sim 0.1$\,mV/K.} 
%
Our maximum value also exceeds the experimental data on ion-exchange membrane systems reviewed in  Ref.~\citenum{barragan2018perspectives} by one order of magnitude. 
\red{Note that the large thermoelectric effects observed here are specific to the nanoscale. Indeed, as detailed in Ref. \cite{sm}, 
$M_{\rm te}$ and $ZT$ should decrease with the channel size $d$, and vanish in macroscopic pores.} 

However, there is still room to enhance the thermoelectric performance of nanofluidic devices, by addressing several key factors appearing in $ZT$. First, we have shown that a large slip length $b$ results in high EO mobility and thermoelectric coefficient, so high-slip surfaces are preferred.  
Second, in contrast with the traditional solid thermoelectric materials, a small electrical conductivity is favorable to enhance $ZT$, for a given $M_{\rm te}$. Since thermoelectric transport arises from the counter-ions at the interface, the only way to reduce the electric conductivity is by reducing the ion concentration in the bulk.
To assess that point, we conducted an additional set of simulations on a \red{salt-free} system containing only counter-ions, with the most hydrophobic and homogeneously charged surface tested above. For $\Sigma = -30$\,mC/m$^2$, we obtained a high $ZT$ of ca. $2.7$, comparable to that of the best performing room temperature thermoelectric materials such as nanostructured Bi$_2$Te$_3$ and Bi$_2$Se$_3$. 
Of course this giant $ZT$ was obtained for a somewhat ideal system, with homogeneous surface charge and large slip length. However, in real systems where it might be difficult to combine a large, heterogeneous surface charge with a large slip length, we suggest that many other potential levers remain to be explored in order to optimize the heat harvesting performance. 
As a simple example, we have shown that the interfacial enthalpy excess of the solvent plays a key role. One could then add a (neutral) solute to affect the bulk (and interfacial) enthalpy density of water in order to enhance the interfacial enthalpy excess.

\paragraph{Summary}
We \red{computed} 
the thermoelectric coefficient using MD simulations for a model nanofluidic system with electrolyte solutions and charged solid walls. We showed that the standard electrostatic picture of the EDL cannot describe the global thermoelectric transport in nanofluidic systems. First, compared to the ion solvation enthalpy, the electrostatic and osmotic contributions induced by the EDL were found to be negligible. Second, we outlined the dominant role of water molecules enthalpy in the thermoelectric transport of the electrolyte, which is neglected in the standard picture. 
Finally, hydrodynamic slip can largely enhance the thermoelectric coefficient, and should be taken into account in the modeling and engineering of such transport process. 
In particular, we showed that the spatial distribution of the surface charge has a strong impact on slip, and hence on the thermoelectric coefficient. 
\red{Better performance was obtained for a homogeneous surface charge, representative of e.g. polarized surfaces.} 
We also investigated the heat harvesting efficiency displayed by these nanofluidic systems by computing the so-called figure of merit and the Seebeck coefficient. 
\red{We discussed the interest of reducing the salt concentration, and} 
we showed that figures of merit comparable to those of state-of-the-art solid-state thermoelectric materials could be obtained \red{in the salt-free limit}. Although our simple model neglects practical effects that could limit the performance of experimental systems, we hope our results will motivate further theoretical and experimental work toward the realization of efficient nanofluidic waste heat harvesters.

\begin{acknowledgments}
The authors thank A.-L. Biance, C. Ybert and C. Cottin-Bizonne for fruitful discussions. 
This work is supported by the ANR, Project ANR-16-CE06-0004-01
NECtAR. LJ is supported by the Institut Universitaire de France. SM acknowledges  support from the H2020 programme FET-open project EFINED (project number 766853). 
\end{acknowledgments}


%

\end{document}